\documentclass[prd,aps,showpacs,tightenlines,superscriptaddress,amsmath,amssymb,amsfonts, nofootinbib]{revtex4-2}
\usepackage{amssymb,amsmath,amsthm,amsbsy,epsfig,color,graphicx,times,physics}
\usepackage{subcaption}
\usepackage[ansinew]{inputenc}
\usepackage[english]{babel}
\usepackage{color}
\usepackage{braket}
\usepackage{tabularx}
\usepackage{array}
\usepackage{slashed}
\usepackage{tikz}
\usetikzlibrary{arrows.meta, positioning}
\usepackage{lmodern}
\usepackage{float}

\begin{document}
	
	\title{Particle mixing and quantum reference frames}
	
	\author{A. Capolupo}
	\email{capolupo@sa.infn.it}
	
	\author{G.Pisacane}
	\email{gpisacane@unisa.it}
	\affiliation{Dipartimento di Fisica ``E.R. Caianiello'' Universit\'a di Salerno,
		and INFN -- Gruppo Collegato di Salerno, Via Giovanni Paolo II, 132,
		84084 Fisciano (SA), Italy}
	
	\author{A. Quaranta}
	\email{aniello.quaranta@unicam.it}
	\affiliation{School of Science and Technology, University of Camerino, Via Madonna delle Carceri, Camerino, 62032, Italy}
	
	\begin{abstract}
		We discuss the problem of defining rest frames for mixed particles, showing that Quantum Reference Frames are necessary to incorporate mass superpositions. This approach reveals a strictly frame-dependent nature of entanglement. We investigate the phenomenological impact on neutrinos and neutral mesons, demonstrating that the transition to the rest frame generates entanglement {\cite{MainPaper}.}
	\end{abstract}
	
	\maketitle
	
	\section{Introduction}
	
	The idea that reference frames may possess quantum features first appeared in quantum information {theory} 
	~\cite{QI1,QI2,QI3,QI4,QI5,QI6}, where \emph{{Quantum Reference Frames} 
	} (QRFs) have first emerged.
	A QRF is a reference frame that is described using the principles of quantum theory. As with any reference frame, it serves as an abstract coordinate system used to define physical quantities like time, position, momentum, and spin. However, since it is formulated within the framework of quantum mechanics, it exhibits several distinctive features and symmetries that do not arise in a conventional classical reference frame.
	
	Given their close link to the relational paradigm and the issue of relational observables~\cite{QI6,RQM}, it was soon realized that QRFs could be particularly significant in the context of Quantum Gravity~\cite{QG1,QG2,QG3,QG4}. This prompted a new, foundational approach~\cite{RF1,RF2,RF3,RF4}, in which QRFs are introduced independently of information-theoretic considerations.
	
	The basic idea behind QRFs leverages on the physical notion of a reference frame, namely that of a \emph{{concrete}} physical system to serve as a reference for the description of physics, rather than the corresponding abstraction. Since it is widely believed that physical systems are ultimately quantum in nature, it is immediate that physical reference frames themselves should be considered quantum. This simple idea has profound consequences. First, the set of transformations connecting the physical frames is naturally extended to incorporate quantum features (e.g., the superposition principle), so that classical transformations such as boosts, translations, and rotations acquire an operator-theoretic nature. Secondly, once quantum superpositions are allowed to play the role of physical reference frames, quantum entanglement becomes a frame-dependent phenomenon.
	QRFs seem to follow a natural historical path of generalization of the concept of \emph{{frame}} and \emph{{frame transformations}}, much like special and general relativity have upgraded the Galilean-Newtonian conceptions.
	
	The key importance of the symmetry group governing transformations between reference frames was clarified by Wigner~\cite{Wigner}. In particular, particle properties are tightly connected to the representations of the Poincar\'{e} group in special relativity. This paper reports on the findings of~\cite{MainPaper}, in which the role of QRFs in particle physics, specifically pertaining to particle mixing, is elucidated. In particular, we demonstrate that mixed particle states can only be fully understood by employing quantum reference frames.
	Indeed, we show that the notion of a rest frame for a mixed particle can emerge solely through a genuinely quantum transformation between frames.
	
	In this context, we explore the frame-dependence of entanglement and discuss how this relativity of entanglement may play a role in processes involving mixed particles. We extend the Wigner framework~\cite{Wigner} in order to incorporate quantum frame transformations and properly describe mixed elementary particles, such as neutrinos. Regarding flavor oscillations, introducing quantum reference frames does not modify the transition probabilities.
	
	The paper is organized as follows: in Section \ref{sec2}, we show the necessity of quantum reference frames to define a rest frame for mixed states. Section \ref{sec3} examines a major implication of quantum frames: the relativity of entanglement, and how this frame-dependent entanglement may appear in decays of mixed particles, and discusses potential observable effects associated with frame-related entanglement. Finally, Section \ref{sec4} summarizes our conclusions.

	\section{Rest Frame of Mixed Particles}\label{sec2}
	
	Most of the elementary particles belong to some sharply defined irreducible projective representation (irrep) of the Poincar\`{e} group. As it is well-known the irreps of the Poincar\`{e} group are fully specified by the eigenvalues of the Casimir operators $P^\mu P_\mu$ and $W^\mu W_\mu$. Massive particle states of an irreps $\lbrace m,s\rbrace$ are labeled by mass $P^\mu P_\mu \ket{m,s}=m^2\ket{m,s}$ and spin $W^{\mu}W_\mu \ket{m,s}=-m^2s(s+1)$. For most applications, the momentum basis of $1$-particle states is the most convenient:
	\begin{equation*}
		P^\mu \ket{k^\mu,\sigma}_{m,s} = k^\mu \ket{k^\mu,\sigma}_{m,s}
	\end{equation*}
	with $\sigma = -s,-s+1,...,s$ denoting a spin component. Here we use the somewhat redundant ${m,s}$ subscript to underline that $\ket{k^\mu,\sigma}_{m,s}$ belongs to the irrep with mass $m$ and spin $s$. The unitary action of the Poincar\`{e} group on these states is given explicitly as~\cite{Wigner,Weinberg}\vspace{-12pt}
	\begin{eqnarray*}
		\\ U(\Lambda) \ket{k^\mu,\sigma}_{m,s} &=& \sum_{\sigma'} D^{(s)}_{\sigma,\sigma'}(W(\Lambda,k)) \ket{\Lambda^\mu_\nu k^\nu,\sigma'}_{m,s} \\ U(a) \ket{k^\mu,\sigma}_{m,s} &=& e^{-ia^\mu k_\mu}\ket{k^\mu,\sigma}_{m,s}
	\end{eqnarray*}
	where $\Lambda$ denotes a (homogeneous) Lorentz transformation, $W(\Lambda,k)$ is the associated Wigner rotation, and $D^{(s)}_{\sigma, \sigma'} (R)$ is the $2s+1$-dimensional matrix representation of the rotation group. Notice, crucially, that the orbits are all contained within a single, sharply defined, irrep.
	Fundamental particles are assumed to carry a well-defined mass and spin. For $m>0$ it is always possible to define the rest frame of a given particle as the frame in which the particle has vanishing spatial momentum. Letting $\ket{\left(E (\pmb{k}) , \pmb{k}, \sigma \right) }_{m,s}$ denote the generic $4$-momentum state, with $\pmb{P} \ket{\left(E (\pmb{k}) , \pmb{k}, \sigma \right)}_{m,s} = \pmb{k} \ket{\left(E (\pmb{k}) , \pmb{k}, \sigma \right)}_{m,s}$, $P^0\ket{\left(E (\pmb{k}) , \pmb{k}, \sigma \right)}_{m,s} = E(\pmb{k}) \ket{\left(E (\pmb{k}) , \pmb{k}, \sigma \right)}_{m,s}$ and $E(\pmb{k}) = \sqrt{m^2 + \pmb{k}^2}$, the state of the particle in its rest frame has, by definition, the form $\ket{(m,\pmb{0},\sigma)}_{m,s}$.

	\subsection{A First Glance at QRFs: Momentum Superpositions}
	
	Let us now consider a particle with definite mass and spin characterized, within some frame, by a momentum superposition (e.g., a wave packet). To simplify the description as much as possible, take it as a superposition of two momenta $\pmb{k}, \pmb{k}'$. Since $m$ and $s$ are unambiguous, and we assume the superposition to be characterized by a common spin state $\sigma$, we omit all these labels and write the state as follows:
	\begin{equation*}
		\ket{\psi} = \cos \alpha \ket{\left(E(\pmb{k}), \pmb{k} \right)} + \sin \alpha \ket{\left(E(\pmb{k}'), \pmb{k}'\right)} \ .
	\end{equation*}
	{If} 
	the particle was characterized by a sharp momentum $k^\mu$, the transformation to its reference frame would be unambiguously defined by requiring $\Lambda^\mu_\nu k^\nu \equiv (m,\pmb{0}) $, i.e., by selecting those Lorentz transformations $\Lambda^\mu_\nu (k)$ that bring $k^\mu$ to its rest form. But how do we transform the superposition state to its rest form?
	A direct scrutiny reveals that a single classical Lorentz transformation cannot simultaneously annihilate $\pmb{k}$ and $\pmb{k}'$. The only possibility is to \emph{{upgrade} 
	} our definition of Lorentz transformations to accommodate the superposition principle, namely by endowing the transformation parameter (the classical momentum vector $k$) with a quantum operatorial nature, switching to a quantum form of Lorentz transformations $\Lambda^\mu_\nu (\pmb{P})$. When $\Lambda^\mu_\nu (\pmb{P})$ acts on a sharp momentum state, it simply performs the corresponding \emph{classical} transformation to the rest form:
	\begin{equation*}
		U(\Lambda(\pmb{P})) \ket{\left(E(\pmb{k}),\pmb{k}\right)} = U(\Lambda(\pmb{k})) \ket{\left(E(\pmb{k}),\pmb{k}\right)} = \ket{\left(m,\pmb{0}\right)} \ .
	\end{equation*}
	{As} a consequence, its action on the superposition simply gives the following:
	\begin{equation*}
		U(\Lambda(\pmb{P}))\ket{\psi} = \left(\cos \alpha + \sin \alpha\right) \ket{(m,\pmb{0})} \ .
	\end{equation*}
	{Since} the superposition occurs within the same irrep, the equality above is not quite correct. To ensure unitarity, the operator would need to include a normalization factor $\frac{1}{\cos \alpha + \sin \alpha}$ so as to finally obtain $ U(\Lambda(\pmb{P}))\ket{\psi} = \ket{(m,\pmb{0})}$. We shall see that this problem does not arise when distinct irreps are involved in the superposition. This apparent conundrum regarding the normalization factor is resolved by considering the particle described by the state $\ket{\psi}$ in a given frame, say $A$, as the (quantum) reference frame to which we wish to transform. To be as precise as possible, we employ the notation $\ket{\psi}_p^{(A)}$, in which the frame $A$ and the particle $p$ to which the state refers are denoted explicitly. Let the reference $A$ be another quantum system, characterized by whatever internal quantum numbers and by an overall three-momentum $\pmb{K}_A$. In the rest frame of $A$, of course, $\pmb{K}_A = 0$. According to $A$, the state of the system $A + p$ is therefore as follows:
	\begin{equation*}
		\ket{\psi}_{Ap}^{(A)} = \ket{\pmb{0}}^{(A)}_{A} \ket{\psi}_p^{(A)} = \ket{\pmb{0}}^{(A)}_{A} \left(\cos \alpha \ket{\left(E(\pmb{k}), \pmb{k} \right)}_p^{(A)} + \sin \alpha \ket{\left(E(\pmb{k}'), \pmb{k}'\right)}_p^{(A)} \right) \ .
	\end{equation*}
	{Such} state lives in the Hilbert space $\mathcal{H}^{(A)}_A \otimes \mathcal{H}^{(A)}_p$ for systems $A$ and $p$ with respect to the frame $A$. Notice that the superposition state is re-obtained by tracing with respect to the frame degrees of freedom $\ket{\psi}=\mathrm{Tr}_A\ket{\psi}_{Ap}^{(A)}$. The QRF transformation to the rest frame of $p$ is the unitary map $U(\Lambda(\pmb{P})): \mathcal{H}^{(A)}_A \otimes \mathcal{H}_p^{(A)} \rightarrow \mathcal{H}^{(p)}_A \otimes \mathcal{H}^{(p)}_{p}$ that takes the momentum of $p$ in the frame $A$ as input and performs the corresponding transformation to the rest frame on \emph{both $A$ and $p$}:
	\begin{equation*}
		U(\Lambda(\pmb{P}))\ket{\psi}_{Ap}^{(A)} = \ket{\psi}_{Ap}^{(p)} = \left( \cos \alpha\ket{\Lambda^\mu_\nu(\pmb{k})K^\nu_A}_A^{(p)} +\sin \alpha\ket{\Lambda^\mu_\nu(\pmb{k}')K^\nu_A}_A^{(p)}\right) \ket{(m,\pmb{0})}_p^{(p)}
	\end{equation*}
	where $K_A^\nu \equiv (M_A, \pmb{0})$ is the total $A$ momentum in its rest form. Notice, importantly, that the map is now automatically unitary, with no need to impose a normalization factor, having the $p$ superposition been transformed into an equivalent $A$ superposition. The apparent problem discussed before is now understood to actually originate from a hidden trace operation $\ket{\psi}_p^{(p)} = \mathrm{Tr}_A \ket{\psi}_{Ap}^{(p)} = \left(\cos \alpha + \sin \alpha \right) \ket{(m,0)}_p^{(p)}$.

	The trivial observations above showcase the need for \emph{quantum} frame transformations to make sense of the rest frame of a quantum superposition of momenta.
	
	\subsection{Mass Mixing}
	
	Up to now our considerations have been limited to a single irrep of the Poincar\`{e} group. All fundamental particles but one do indeed carry a definite mass. The notable exception is represented by flavor neutrinos~\cite{N1,N2,N3,N4,N6,N7,N8,N9,N10,N13,N14}. They are considered as elementary particles, but at the same time, they do not, at least strictly speaking, belong to an irrep $\ket{m,s}$. We shall drop the spin label from now on, since it plays no role in the following arguments. We shall also limit ourselves to two-flavor mixing, even though none of our considerations depend critically on the number of flavors.

	The state of an electron neutrino can be written as the superposition:
	\begin{equation}\label{NeutrinoState}
		\ket{\pmb{k}, e} = \cos \theta \ket{k^{\mu}_{1}}_{m_1} + \sin \theta \ket{k_{2}^{\mu}}_{m_2}
	\end{equation}
	{The} momenta are specified as $k_{j}^{\mu} \equiv \left(E_j( \pmb{k}), \pmb{k} \right)$ with the on-shell energies $E_j(\pmb{k}) = \sqrt{\pmb{k}^2 + m_j^2}$.
	The superposition \eqref{NeutrinoState} is still an eigenstate of the $3$-momentum operator $\pmb{P}$, but does not possess a definite energy. Muon neutrino states are completely analogous. Each transformation that involves the time direction acts non-trivially on the state \eqref{NeutrinoState}, leading, among other things, to the phenomenon of neutrino oscillations. At least $N_f$ distinct irreps are necessary to construct neutrino flavor states, with $N_f$ the number of flavors.
	
	From the previous subsection, we may expect that the rest frame of a flavor neutrino does only make sense as a QRF, since a quantum superposition is unavoidable. It is immediate to verify that this is the case, since there does not exist a classical Lorentz boost able to simultaneously annihilate the $3$-momentum of the $m_1$ and $m_2$ components of the neutrino state of Equation \eqref{NeutrinoState}. To identify the QRF transformation needed to reach the neutrino rest frame, first promote the Lorentz boost parameter to an operator
	\begin{equation}
		\hat{\psi} = \cosh^{-1} \left( \hat{H}\left( \hat{P}^{\mu} \hat{P}_{\mu} \right)^{-\frac{1}{2}} \right) \ ,
	\end{equation}
	with $\hat{H} = \hat{P}^0$, and the hats have been introduced to underline the operatorial nature. This operatorial definition extends the classical boost parameter $\psi = \cosh^{-1} \left(\frac{E}{m} \right)$. Assuming $\pmb{k}$ oriented along the $z$ axis, without loss of generality, the quantum Lorentz boost is as follows:
	\begin{equation}\label{QuantumTransform}
		\hat{\Lambda}^{\mu}_{\nu} \equiv \begin{pmatrix} \hat{H}\left( \hat{P}^{\mu} \hat{P}_{\mu} \right)^{-\frac{1}{2}} & 0 & 0 & -\hat{P}_z\left( \hat{P}^{\mu} \hat{P}_{\mu} \right)^{-\frac{1}{2}} \\ 0 & 1 & 0 & 0 \\ 0 & 0 & 1 & 0 \\ -\hat{P}_z\left( \hat{P}^{\mu} \hat{P}_{\mu} \right)^{-\frac{1}{2}} & 0 & 0 & \hat{H}\left( \hat{P}^{\mu} \hat{P}_{\mu} \right)^{-\frac{1}{2}} \end{pmatrix} \
	\end{equation}
	{The} formal consistency of the quantum boost $\hat{\Lambda}_{\nu}^{\mu}$ is intrinsically linked to the algebraic properties of the Poincaré generators. In standard Quantum Field Theory, the components of the four-momentum operator satisfy the commutation relations $[\hat{P}^{\mu}, \hat{P}^{\nu}] = 0$. This implies that the Hamiltonian $\hat{H}$ and the squared mass operator $\hat{M}^2 = \hat{P}^{\mu}\hat{P}_{\mu}$ are mutually commuting self-adjoint operators. Consequently, any operator-valued function defining the boost, such as the parameter $\hat{\psi}$ in Equation (2), is uniquely defined through functional calculus for commuting operators, avoiding any operator-ordering ambiguities. Furthermore, by restricting the analysis to the massive sector of the Hilbert space ($\mathcal{H} = \bigoplus_{m>0} \mathcal{H}_m$), the operator $(\hat{M}^2)^{-1/2}$ remains bounded and well-defined on the domain of physical states, ensuring that the transformation is a non-singular unitary representation.
	The associated unitary representation $U(\hat{\Lambda})$ transforms the neutrino state of Equation \eqref{NeutrinoState} into its rest form, simultaneously annihilating the $3$-momentum of the $m_1$ and $m_2$ components.
	
	Although we have focused exclusively on the rest frame, QRF transformations are needed \emph{{whenever we wish to move to a different frame of reference and maintain a definite neutrino $3$-momentum}}. For instance, shifting to a frame in which the neutrino momentum is $\pmb{k}' \neq \pmb{k}$, still oriented along $z$, requires a QRF transformation with a boost parameter
	\begin{equation}
		\hat {\psi} = \sinh^{-1} \left( \left(\hat{P}^{\mu} \hat{P}_{\mu}\right)^{-1} \left(k' \hat{H} - \sqrt{k^{'2} + \hat{P}^{\mu} \hat{P}_{\mu}} \hat{P}_z \right) \right) \ .
	\end{equation}
	{We} can see that QRFs and QRF transformations are unavoidable to properly associate reference frames to mixed particles.
	
	It is legitimate to claim that defining the “rest frame” of a particle with an indefinite four-momentum, such as a wave packet
	$\ket{\phi} = \int d^4 p \ \theta (p^0) \ \delta (p^{\mu} p_{\mu} - m^2) \phi (p) \ket{p^{\mu}}_{m}$,
	requires a quantum version of the Lorentz transformation. This is indeed correct: whenever states involve superpositions of momenta, quantum transformations become necessary.
	
	However, neutrino flavor states differ fundamentally from standard wave packets, as quantum transformations play a critical role in defining their associated rest frame. First, while ordinary wave packets are typically formed by superposing states on the same mass shell, flavor states inherently involve at least two different masses. Second, unlike elementary particles, which can occupy states with definite four-momentum, flavor states are intrinsic superpositions of different four-momenta, possessing an unavoidable, built-in uncertainty.
	
	It is crucial to emphasize that this QRF transformation cannot be reduced to the elementary notion of a mass-dependent “controlled boost”. While a standard controlled operation simply transforms a system based on its internal mass eigenvalues, the QRF map represents a fundamental shift in perspective between relational Hilbert spaces. By incorporating a generalized “parity-swap” mechanism, the transformation effectively redistributes the quantum fluctuations—and the inherent superpositions—of the reference frame onto the observed systems. This redistribution is the conceptual origin of the relativity of entanglement: the “quantumness” previously attributed to the frame is shifted onto the relational state of the particles, as explored in detail in the following section.

	\section{Relativity of Entanglement for Mixed Particles}\label{sec3}
	
	The arguments of the previous section hold for any kind of mixed particle, including fundamental (neutrinos) and composite, such as neutral mesons~\cite{Meson1,Meson2,Meson5,Meson7,Meson10,Meson11,Meson12}, and the mixing of neutrons and e.g., mirror neutrons~\cite{Mirror1,Mirror2,Mirror3}. The introduction of QRFs leads to \emph{{relativity of entanglement}}. Quantum entanglement turns out to be a strongly frame-dependent concept. To illustrate this concept, consider two quantum particles $A,B$ and an electron neutrino $\nu$. Let the state of $B$ and $\nu$, with respect to $A$, be as follows:
	\begin{equation}\label{Astate}
		\ket{\psi_{B\nu}^{(A)}} = \ket{k_B^{\mu}}_B^{(A)} \left( \cos \theta \ket{k^{\mu}_{1}}_{m_1, \nu}^{(A)} + \sin \theta \ket{k_{2}^{\mu}}^{(A)}_{m_2, \nu} \right) \ .
	\end{equation}
	{The} state of $A$ with respect to $A$, i.e., its rest state $\ket{k^{\mu}_{A,0}}_A^{(A)} = \ket{(m_A,\pmb{0})}_A^{(A)} $, is omitted for notational convenience. Let us now transform to the rest frame of $\nu$ via the Quantum Lorentz boost \eqref{QuantumTransform}. The state of $A,B$ with respect to $\nu$ is as follows:
	\begin{eqnarray}\label{NeutrinoEntangledState}
		\ket{\psi_{AB}^{(\nu)}} = \cos \theta \ket{\Lambda^{\mu}_{1,\nu}p_{A,0}^{\nu}}_A^{(\nu)} \ket{\Lambda^{\mu}_{1,\nu}p_{B}^{\nu}}_B^{(\nu)}
		+\sin \theta \ket{\Lambda^{\mu}_{2,\nu}p_{A,0}^{\nu}}_A^{(\nu)} \ket{\Lambda^{\mu}_{2,\nu}p_{B}^{\nu}}_B^{(\nu)} \ ,
	\end{eqnarray}
	where $\Lambda_j$ is the classical boost taking the $4$-momentum $\left(E_j(\pmb{k}),\pmb{k} \right)$ to its rest form. In contrast to the initial state \eqref{Astate}, which is factorizable, the transformed state of Equation \eqref{NeutrinoEntangledState} is manifestly entangled. In other words, the QRF transformation has turned the quantum superposition characterizing $\nu$ into an entanglement between $A$ and $B$.
	
	\subsection{Entanglement in Neutral Meson Decays}
	To show how the frame related entanglement may impact actual physical processes, we focus, for concreteness, on the decay of a charged $D$ meson into three particles~\cite{PDG} $D^{+} \rightarrow \bar{K}^0 + e^+ + \nu_e $.

	The strangeness eigenstate (CP violation is neglected for simplicity) $\ket{\bar{K}^0} = \frac{1}{\sqrt{2}}\left(\ket{K_L} - \ket{K_S} \right)$ is a linear combination of the mass eigenstates $\ket{K_L}, \ket{K_S}$, which have definite values of mass $m_L, m_S$. Assuming that the Kaon has definite $3$-momentum $\pmb{k}$ in the laboratory frame, we can write the state of the decay products as follows:
	\begin{eqnarray}
		\nonumber &&\ket{DP}^{LAB} = \frac{1}{\sqrt{2}}\left(\ket{(E_L(\pmb{k}),\pmb{k}}_{m_L, K}^{LAB} - \ket{(E_S(\pmb{k}),\pmb{k}}_{m_S, K}^{LAB} \right)\ket{k^{\mu}_{e^+}}_{e^+}^{LAB} \ket{k^{\mu}_{\nu_e}}_{\nu_e}^{LAB} \ .
	\end{eqnarray}
	{The} same process, characterized by the same kinematical invariants, may be described also in the rest frame of the Kaon. The latter may be identified with the laboratory frame if the Kaon is nearly at rest. The state of the decay products in the Kaon frame is obtained once again via a QRF transformation:
	\begin{eqnarray}\label{EntangledKaon}
		\nonumber \ket{DP}^{(K)} &=& \frac{\ket{\Lambda^{\mu}_{L,\nu}k^{\nu}_{e^+}}_{e^+}^{(K)} \ket{\Lambda^{\mu}_{L,\nu}k^{\nu}_{\nu_e}}_{\nu_e}^{(K)}\ket{(m_L,\pmb{0})}_{m_L, K}^{(K)} }{\sqrt{2}} \ \\ &-& \frac{\ket{\Lambda^{\mu}_{S,\nu}k^{\nu}_{e^+}}_{e^+}^{(K)} \ket{\Lambda^{\mu}_{S,\nu}k^{\nu}_{\nu_e}}_{\nu_e}^{(K)}\ket{(m_S,\pmb{0})}_{m_S, K}^{(K)}}{\sqrt{2}} \ .
	\end{eqnarray}
	{Just} like the QRF transformation to the neutrino rest frame induces the entanglement of the state \eqref{NeutrinoEntangledState}, the transformation to the rest frame of the Kaon induces entanglement among the leptons in the state \eqref{EntangledKaon}.
	Remarkably, this form of entanglement is exclusively related to QRF transformations.

	\subsection{Observable Signatures}\label{3.2}
	The appearance and disappearance of entanglement, based solely on the choice of a frame, appears rather surprising, given that all the transformations involved are unitary by definition~\cite{RF7}.
	To clarify such point, it is useful to think in terms of concrete Entanglement-related observables. The latter are simply mapped (unitarily) into distinct observables by QRF transformations, so there is no 'loss of information' whatsoever. Consider, for instance, the
	states of Equations \eqref{Astate} and \eqref{NeutrinoEntangledState}. The quantum boost transforming from the $A$ frame to the $\nu$ frame is the unitary map
	\begin{equation}
		U(\Lambda(\pmb{P})) : \mathcal{H}^{(A)}_B \otimes \mathcal{H}^{(A)}_{\nu} \rightarrow \mathcal{H}_A^{(\nu)} \otimes \mathcal{H}_B^{(\nu)}
	\end{equation}
	that takes the $\nu$ momentum as input and performs the corresponding classical boost on $A$ and $B$. If $O^{(\nu)}_{AB} : \mathcal{H}_A^{(\nu)} \otimes \mathcal{H}_B^{(\nu)} \rightarrow \mathcal{H}_A^{(\nu)} \otimes \mathcal{H}_B^{(\nu)}$ is an observable in the $\nu$ rest frame, the inverse QRF transformation $U^{\dagger}$ maps $O$ unitarily to an observable in the $A$ frame as~\cite{RF2}:
	\begin{equation}
		O^{(A)}_{B\nu} = U^{\dagger}O^{(\nu)}_{AB}U \ .
	\end{equation}
	{Here} we have dropped the argument of $U$ for compactness. As a consequence of unitarity, one has
	\begin{equation}\label{Unitarity}
		\bra{\psi^{(A)}_{B\nu}} O^{(A)}_{B\nu} \ket{\psi^{(A)}_{B\nu}} = \bra{\psi^{(\nu)}_{AB}} O^{(\nu)}_{AB} \ket{\psi^{(\nu)}_{AB}} \ ,
	\end{equation}
	or, equivalently, $\mathrm{Tr}_{AB}(\rho^{(\nu)}_{AB}O^{(\nu)}_{AB}) = \mathrm{Tr}_{B\nu}(\rho^{(A)}_{B\nu}O^{(A)}_{B\nu})$, in terms of the density matrices $\rho^{(\nu)}_{AB}$ and $\rho^{(A)}_{B\nu}$.
	While the two sides of Equation \eqref{Unitarity} have the same numerical value, they have wildly different interpretations in the two frames, to the point that they disagree on the (reduced) system with respect to which the observables are defined {(Of course, the \emph{global system} is $AB\nu$ for both.)} 
	($B\nu$ or $AB$). Quantum correlations that manifest as entanglement in the neutrino frame are unitarily mapped into non-entanglement-related observables in frame $A$.
	We can make this point even clearer by defining an appropriate entanglement witness (i.e., an observable signalling the presence of entanglement) on the state \eqref{NeutrinoEntangledState}.
	We set $\bar{k}^{\mu}_{A,1}=\Lambda^{\mu}_{1,\nu}k^{\nu}_{A,0}, \bar{k}^{\mu}_{A,2}=\Lambda^{\mu}_{2,\nu}k^{\nu}_{A,0}, \bar{k}^{\mu}_{B,1}=\Lambda^{\mu}_{1,\nu}k^{\nu}_{B}, \bar{k}^{\mu}_{B,2}=\Lambda^{\mu}_{2,\nu}k^{\nu}_{B}$, and rewrite the state \eqref{NeutrinoEntangledState} as follows:
	\begin{equation}
		\ket{\psi^{(\nu)}}_{AB} = \cos \theta \ket{\bar{k}^{\mu}_{A,1}}_A^{(\nu)} \ket{\bar{k}^{\mu}_{B,1}}_B^{(\nu)} + \sin \theta \ket{\bar{k}^{\mu}_{A,2}}_A^{(\nu)} \ket{\bar{k}^{\mu}_{B,2}}_B^{(\nu)} \ .
	\end{equation}
	{We} define a tensor product operator $W_{AB}^{(\nu)}$ as
	\begin{equation}
		W_{AB}^{(\nu)} = W_A \otimes W_B : \mathcal{H}_A^{(\nu)} \otimes \mathcal{H}_B^{(\nu)} \rightarrow \mathcal{H}_A^{(\nu)} \otimes \mathcal{H}_B^{(\nu)}
	\end{equation}, with each of the $W_J$ having as the only non-vanishing elements
	\begin{equation}
		\bra{\bar{k}^{\mu}_{J,2}} W_{J} \ket{\bar{k}^{\mu}_{J,1}} = w_{J}
	\end{equation}
	for $J = A,B$ and some complex parameters $w_{J}$. One immediately checks that $W_{AB}^{(\nu)}$ acts as an entanglement witness. The expectation value of $W_{AB}^{\nu}$ vanishes on factorizable states. On the contrary,
	\begin{equation}\label{Witness}
		\bra{\psi_{AB}^{(\nu)}} W^{(\nu)}_{AB} \ket{\psi_{AB}^{(\nu)}} = 2 \cos\theta \sin \theta \ \mathrm{Re} \left(w_A w_B \right) \ .
	\end{equation}
	{The} observable $W$ is proportional to the linear entropy of the state
	\begin{equation}\label{E}
		E_L = \frac{\bra{\psi_{AB}^{(\nu)}} W^{(\nu)}_{AB} \ket{\psi_{AB}^{(\nu)}}^2}{2 \mathrm{Re}^2 \left( w_A w_B\right)} \ ,
	\end{equation}
	showing clearly the role of $W$ as a proper witness in the $\nu$ frame. In the $A$ frame, on the other hand, \eqref{Witness} is transformed into the expectation value
	\begin{equation}
		\bra{\psi^{(A)}_{B\nu}} W^{(A)}_{B\nu} \ket{\psi^{(A)}_{B\nu}} = 2 \cos \theta \sin \theta \mathrm{Re} \left(w_A w_B \right) \ ,
	\end{equation}
	which shows no relation to entanglement at all (recall that \eqref{Astate} is factorizable). Therefore, it is possible to indirectly probe frame-related entanglement, even in a frame where entanglement is absent, just by measuring the suitable observable.
	To assess the phenomenological relevance of the frame-dependent entanglement, we provide order-of-magnitude estimates for the linear entropy $E_L$. For a bipartite qubit-like system, the linear entropy serves as a measure of entanglement ranging from $0$ (separable state) to a maximum value of $0.5$ (maximally entangled state). As derived from Equation \eqref{E}, the degree of entanglement generated by the QRF transformation is governed by the mixing angle $\theta$ according to the following:
	\begin{equation}
		E_L = 2 \cos^2 \theta \sin^2 \theta = \frac{1}{2} \sin^2(2\theta).
	\end{equation}
	{By} considering established experimental values for particle mixing, we can quantify this effect:
	\begin{itemize}
		\item Solar Neutrinos: Taking the experimental mixing angle $\theta_{12} \approx 34^\circ$, the resulting linear entropy in the neutrino rest frame is $E_L \approx 0.43$.
		\item Neutral Kaons ($K^0 - \bar{K}^0$): In this system, the mixing is nearly maximal ($\theta \approx 45^\circ$), leading to $E_L \approx 0.5$.
	\end{itemize}
	{These} values demonstrate that the entanglement induced by the QRF transformation is not a negligible correction, but is in fact \textit{{nearly maximal} 
	} for both neutrinos and neutral mesons. This indicates that the transition from the laboratory frame to the particle's rest frame fundamentally alters the system's quantum correlation structure, turning a mass-state superposition into a dominant entanglement signature that, in principle, is accessible to experimental verification.
	
	\section{Conclusions}\label{sec4}
	Following a brief historical overview of QRFs, we discussed the motivations behind their introduction and their relevance in relational approaches. We reported on the findings of~\cite{MainPaper}, where QRFs are recognized as essential to the conception of reference frames associated with mixed particles. The phenomenological relevance of this 'relativity of entanglement' is underscored by the possibility of detecting frame-induced signatures in accelerator-based experiments. Specifically, the decay of mesons, such as $D^+$ or $B^+$, provides a concrete arena to test these predictions. The entanglement witness $W$ introduced in Section \ref{3.2} can be unitarily mapped to laboratory observables, allowing researchers to probe the quantum correlations of the rest frame directly from the laboratory perspective. Future experiments at high-luminosity facilities, such as Belle II or LHCb, could potentially identify these signatures by analyzing the kinematic distributions of decay products with unprecedented precision. Beyond particle physics, the use of QRFs for mixed states has profound implications for foundational issues in Quantum Gravity. The relational paradigm advocated in this work suggests that the very structure of spacetime and the classification of elementary particles should be reformulated to account for the quantum features of observers. This could pave the way for a generalized Quantum Field Theory in which Poincaré symmetry itself acquires an operator-theoretic nature. Such a framework is essential for describing physics in scenarios where gravity and quantum effects are both dominant, providing a bridge between the local description of mixed particles and the global, relational structure of a quantum spacetime.

	\begin{acknowledgments}
		Partial financial support from MIUR and INFN is acknowledged.
		A.C. also acknowledges the COST Action CA1511 Cosmology
		and Astrophysics Network for Theoretical Advances and Training Actions (CANTATA).
	\end{acknowledgments}

\end{document}